\documentclass[10pt,showpacs,twocolumn,nofootinbib,floatfix,aps,prd]{revtex4-1}
\usepackage{amsmath} \usepackage{graphicx} \usepackage{amsfonts}
\usepackage{array} \usepackage{amsthm} \usepackage{bm}
\usepackage{palatino} \usepackage{mathpazo} 
\usepackage{supertabular}

\usepackage[breaklinks]{hyperref}
\usepackage{color}

\usepackage{epstopdf}
\usepackage{graphicx}

\newcommand{\be}{\begin{equation}}
\newcommand{\ee}{\end{equation}}
\newcommand{\ba}{\begin{eqnarray}}
\newcommand{\ea}{\end{eqnarray}}
\newcommand{\bal}{\begin{align}}
\newcommand{\eal}{\end{align}}

\newcommand{\bw}{\begin{widetext}}
\newcommand{\ew}{\end{widetext}}

\begin{document}

\title{\bf \Large Spherical Accretion on a Schwarzschild-MOG Black Hole}

\author{Ayyesha K. Ahmed}\email{ayyesha.kanwal@sns.nust.edu.pk}
\affiliation{Department of Mathematics, Capial University
of Science and Technology (CUST), Islamabad, Pakistan}

\author{M Z A Moughal}\email{zubair.moughal@ceme.nust.edu.pk}
\affiliation{Department of Basic Science and Humanities, Collage of Electrical and Mechanical Engineering, NUST, H-12, Islamabad, Pakistan}

\begin{abstract}
In this paper we have  examined spherical accretion onto Schwarzschild–MOG Black Holes within the framework of Modified Gravity. Using isothermal test fluids, we analyze the behavior of the flow near the critical (sonic) point for various values of the equation of state parameter \( k \). Depending on the fluid type, the flow exhibits either subsonic or supersonic behavior, with ultra-stiff and ultra-relativistic fluids allowing both regimes, while radiation and sub-relativistic fluids show more restricted dynamics. Phase space analysis helps visualize these transitions. We also compute the mass accretion rate and find that it increases with both radial distance and the MOG parameter \( \alpha \), highlighting the role of modified gravity in enhancing accretion processes around black holes.

\end{abstract}



\maketitle

\section{Introduction}\label{INT}
Black holes (BHs) stand as some of the most captivating and enigmatic objects in the universe, embodying the final act in the life cycle of massive stars. The groundbreaking detection of gravitational waves from binary BH mergers, achieved through the Laser Interferometer Gravitational-Wave Observatory (LIGO), has brought their existence into sharp focus \cite{Bp1}. Adding to this evidence, the discovery of star–black hole binary systems using radial velocity measurements offers yet another testament to their profound presence in the cosmos.

Moffat \cite{2} introduced a Modified Gravity (MOG) theory that explains the movement of galaxies and  clusters of galaxies in the present universe without needing dark matter. This theory has also successfully explained observations in the solar system, galaxy rotation curves, and the motion of galactic clusters, the development of cosmic structures, the distribution of matter on large scales, and the acoustic patterns observed in the cosmic microwave background (CMB) spectrum\cite{JWM2, JWM3, JWM4,JWM5, JWM6,JWM7}. J.~W.~Moffat investigated BHs under the framework of Modified Gravity (MOG) \cite{SMOG1}. This theory modifies the gravitational interaction by introducing an effective gravitational constant \( G = G_N (1 + \alpha) \), where \( \alpha \) is a dimensionless parameter and \( G_N \) is Newton’s constant. It also includes a repulsive component characterized by a gravitational charge \( Q = (\alpha G_N)^{1/2} M \), with \( M \) representing the mass of the BH. The possibility of detecting such BHs with the Event Horizon Telescope (EHT) has been explored in \citep{guo2018observational, Moffat:2019uxp}.
 The motion of particles around rotating BHs and BHs in magnetic fields, along with the phenomena of strong and weak gravitational lensing by clusters and supermassive BHs, has been thoroughly explored \citep{sharif2017particle, hussain, moffat2018applying,izmailov2018modified}.

Accretion refers to the process through which a gravitating object gathers surrounding matter, resulting in an increase in its mass and angular momentum \cite{A3}. The accretion of matter onto BHs is a significant phenomenon of enduring interest to astrophysicists. In 1952, Bondi pioneered the study of accretion by developing a model for the spherically symmetric accretion but context of Newtonian physics \cite{Bondi1, Bondi2}.
Later on, Michel \citep{Michel} discussed the BH accretion in the context of GR  and also investigated the accretion onto charged BHs. Shapiro made significant contributions to the understanding of BH accretion by studying the luminosity and frequency spectra of infalling gas \citep{shapiro1973accretion}, the effects of interstellar magnetic fields \citep{shapiro1973magnetic}, and the dynamics of accretion onto rotating BHs \citep{shapiro1974accretion}. The role of gas backreaction in influencing the accretion rate was further analyzed by \cite{malec1999}. Babichev et al. \cite{Babichev} investigated the accretion of dark energy onto a Schwarzschild BH, showing that, unlike normal matter, dark energy accretion can lead to a decrease in BH's mass. Debnath \cite{Debnath} extended this analysis by formulating a general model for static accretion onto static, spherically symmetric BHs. Accretion in higher-dimensional spacetimes has been considered in works such as \citep{giddings2008astrophysical, john2013accretion}. More recently, Das et al. \cite{Das(2024)} studied spherical accretion in higher-dimensional Reissner-Nordström BHs, contributing to a greater understanding of accretion in modified gravity frameworks. Recently, accretion onto BH with backreaction effects on cylindrical symmetrically BHs was discussed \cite{Moughalkamran1}.
\par Accretion BHs in the framework of Scalar-Tensor-Vector Gravity (STVG) was initially explored by \cite{john2016black}, with later developments focusing on accretion disks around MOG BHs \citep{perez2017accretion}. Several recent works have further expanded on accretion dynamics in MOG through analytical and numerical methods \cite{1t, t1, PMach, Mjamil, ahmed, UmarFarooq:2019uqr}. In this study, we investigate spherical accretion onto a Schwarzschild-MOG (SH-MOG) BH, analyzing how various types of fluid behave in this modified gravity background. We derive the SH-MOG metric, formulate the general conservation laws governing the accretion process, and compute critical physical quantities such as the speed of sound at the sonic point. Compared to the classical SH solution in GR, the presence of the MOG parameter introduces deviations in the gravitational potential, which in turn affect accretion rates, sonic point locations, and flow dynamics. Throughout our analysis, we use standard relativistic conventions, adopting the metric signature $(-,+,+,+)$  with geometric units with $ c=G = 1$.
\section{Schwarzschild-MOG Black Hole}
The Schwarzschild-MOG (SH-MOG) BHs represents a static, spherically symmetric vacuum solution to the modified Einstein field equations. The metric in Boyer–Lindquist coordinates $x^\alpha=(t,r,\theta,\phi)$ is given by \cite{SMOG1, JWM8}
\begin{eqnarray}\label{1a}
ds^{2}&=&-f(r)dt^2+\frac{1}{f(r)}dr^{2}+r^2(d\theta^2+\sin^2\theta d\phi^2),
\end{eqnarray}
where,
\begin{eqnarray}\label{2a}
f(r)&=&1-\frac{2(1+\alpha)M}{r}+\frac{\alpha(1+\alpha)M^2}{r^2}.
\end{eqnarray}
Here $M$ denotes the BH's mass, $\alpha$ is the MoG parameter defined  as $\alpha=\frac{G-G_N}{G_N}$ where $G_{N}$ being the Newton's gravitational constant and the parameter  $\alpha$ characterizes the deviation from standard gravity and ranges  $-1\ \text{to} \ \infty$. The horizons of (\ref{1a}) are given as
\begin{eqnarray}\label{3a}
r_{\pm}=M(1+\alpha{\pm}\sqrt{1+\alpha}).
\end{eqnarray}
The ${r_+}$ is being thr event horizon and $r_-$ is considered as the Cauchy horizon. Clearly, $\alpha=0$ reduces it to the Schwarzschild BH. The geodesic equations are given as \cite{HU1}
\begin{eqnarray}
    \Dot{t}&=&\frac{\epsilon}{f},\\
    \Dot{r}&=&\epsilon^2-f(\epsilon-\frac{\mathcal{L}^2}{r^2})\,\\
    \Dot{\phi}&=&\frac{\mathcal{L}^2}{r^2},
\end{eqnarray}
here, $\epsilon$ denotes the specific energy, $\mathcal{L}$ is the angular momentum, and a dot indicates differentiation with respect to an affine parameter $\lambda$. The value 
$\epsilon=0$ corresponds to a null geodesic, while 
$\epsilon=1$ indicates a timelike geodesic.

Curvature invariants can signals the presence of a spacetime singularity. The curvature invariants of SH-MOG BH in (\ref{1a}) are
\begin{eqnarray}
I_{1}&=&g^{\mu\nu}R_{\mu\nu}=0,\label{e1}\\
I_{2}&=&R^{\mu\nu}R_{\mu\nu}=\frac{4M^{4}\alpha^{2}(1+\alpha)^{2}}{r^8},\label{e2}\\
I_{3}&=&R^{\mu\nu\rho\sigma}R_{\mu\nu\rho\sigma}=\quad\frac{8M^{2}(1+\alpha)^{2}}{r^8}\Big(6r^{2}\\\quad  &&~~~~~~~~~~~~~~~~~~~~~~~~~-12Mr\alpha+7M^{2}\alpha^{2}\Big).\label{e3}
\end{eqnarray}
From the above equations, we find that curvature is finite all over the region except $r=0$. Thus at $r=0$, we have a physical singularity and  we have a coordinate singularity at horizon. We can remove the coordinate singularity by transforming the metric into Eddington–Finkelstein coordinates. For this, first we can re-write the metric as
\begin{eqnarray}
\begin{split}
ds^2=-\frac{(r-r_+)(r-r_-)}{r^2}dt^2+\frac{r^2}{(r-r_+)(r-r_-)}dr^2+\\r^2(d\theta^2+\sin^2\theta d\phi^2).
\end{split}
\end{eqnarray}
For radial null geodesics, we get
\begin{equation}
\pm{dt}=\frac{r^2 dr}{(r-r_-)(r-r_+)},
\end{equation}
which on integration gives
\begin{equation} \label{8a}
\pm{t}=t^{\prime}+\ln{\Big|\frac{r}{r_+}-1\Big|\frac{r_+^2}{r_+-r_-}-\ln{\Big|\frac{r}{r_-}-1\Big|}\frac{r_-^2}{r_+-r_-}},
\end{equation}
where $t_-$ is for in-going particles and $t_+$is for outgoing particles. This correspond to the advanced EF and retarted EF coordinates respectively. We consider advance EF coordinate given as
\begin{equation} \label{11a}
dt^{\prime}=dt+dr\Bigg[\frac{(r_++r_-)r+(r_+r_-)}{(r-r_-)(r-r_+)}\Bigg]
\end{equation}
Also Eq. (\ref{3a}) gives
\begin{eqnarray}
r_{-}r_{+}&=&\alpha(1+\alpha)M^2,\label{8b}\\
r_{-}+r_{+}&=&2(1+\alpha)M.\label{9b}
\end{eqnarray}
Substituting \ref{8b} and \ref{9b} in \ref{11a}, we get
\begin{eqnarray}\label{10a}
dt^{\prime}&=&dt+\Big[\frac{(1+\alpha)2Mr+\alpha(1+\alpha)M^2}{r^{2}-(1+\alpha)2Mr+\alpha(1+\alpha)M^2}\Big]dr.
\end{eqnarray}
Hence, the metric becomes:
\begin{eqnarray}\label{met}
\begin{split}
ds^{2}=&&-\Big[1-\frac{(1+\alpha)M}{r}\big(2-\frac{\alpha M}{r}\big)\Big]dt^{\prime2}\\
&&+2\Big[2(1+\alpha)Mr+\alpha(1+\alpha)M^{2}\Big]dt^{\prime}dr\nonumber\\
&&+\Big[1-\frac{(1+\alpha)M}{r}\big(2-\frac{\alpha M}{r}\big)\Big]dr^{2}\\
&&+r^{2}(d\theta^{2}+\sin^{2}\theta d\phi^{2}).
\end{split}
\end{eqnarray}
The above equation (\ref{met}) is in EF Coordinate and coordinate singularity is removed. 
\section{General Equations for Spherical Accretion in MOG}

We begin by formulating the equations that govern spherical accretion, assuming the infalling matter behaves as a perfect fluid. The analysis focuses on both the accretion rate and the flow dynamics, grounded in the conservation laws of particle number and energy. Denoting the particle number density by $n$ and the fluid’s four-velocity by $u^\mu$, the current density is given by
\begin{equation}
    J^{\mu}=n u^{\mu}.
\end{equation}
Based on the principle of particle number conservation, the total number of particles remains constant; in other words, the four-divergence of the current density vanishes. Mathematically, this is expressed as:
\begin{eqnarray}\label{12}
\nabla_{\mu}J^{\mu}&=&0,
\end{eqnarray}
The symbol $\nabla_{\mu}$ denotes the covariant derivative. The energy-momentum tensor for a perfect fluid takes the form given by
\begin{equation}
    T^{\mu \nu}=(e+p)u^{\mu}u^{\nu}+pg^{\mu \nu}
\end{equation} with $e$ tands for the energy density and $p$ represents the pressure. Conservation of energy and momentum is ensured by the vanishing divergence of the energy–momentum tensor,
\begin{eqnarray}\label{13k}
\nabla_{\mu}T^{\mu \nu}&=&0.
\end{eqnarray}
Because Bondi-type accretion is steady and spherically symmetric \cite{1t, t1}, all physical quantities vary only with the radial coordinate $r$. To analyze the fluid’s accretion properties, we employ a Hamiltonian approach. The general analysis for a static, spherically symmetric metric has been presented in \cite{A3}; here, we make use of the following results from that work:
\begin{eqnarray}
    r^2nu^r&=&c_1,\label{13}\\
    h^2[f+(u^r)^2]&=&c_2,\label{13aa}
\end{eqnarray}
where we denoted the particle number density by $n$ and the fluid’s four-velocity by $u^\mu$, $h$ is the specific enthalpy, $f$ is a metric function, and $c_1, c_2$
  are integration constants.
\section{Sonic point}
The sonic point is where the fluid’s inward velocity matches the local speed of sound. At this critical radius, the accretion flow reaches its maximum accretion rate. The local speed of sound at this point is given by
\begin{eqnarray}\label{15}
a^2 &=& \left( \frac{u^r}{u_t} \right)^2,
\end{eqnarray}
where \( a \) denotes the sound speed. If the pressure is assumed to remain constant, the specific enthalpy depends solely on the particle number density, i.e., \( h = h(n) \), indicating that the fluid is barotropic. Under this condition, the equation of state simplifies to
\begin{eqnarray}\label{16}
\frac{dh}{h} &=& a^2 \left( \frac{dn}{n} \right).
\end{eqnarray}
Using normalizing condition $u^\mu u_\mu=-1$, we can also write (\ref{13aa}) as 
\begin{eqnarray}
    h^2(u^t)^2&=&c_2.\label{13aaa}
\end{eqnarray}
Now differentiating  (\ref{13}) and substituing (\ref{13aaa}) in it we get:
\begin{eqnarray}\label{13b}
u^r\frac{du^r}{dr} &=& -\frac{(a u_t)^2}{n}\frac{dn}{dr}-\frac{2(1+\alpha)M}{r^2}\\&+&\frac{\alpha(1+\alpha)M^2}{r^3}.
\end{eqnarray}
Now differentiating (\ref{13aa}), we get
\begin{eqnarray}\label{13ab}
 \frac{1}{n}\frac{dn}{dr}&=& -\frac{2}{r}-u^r\frac{du^r}{dr}
\end{eqnarray}
and substituting in  (\ref{13b}) and simplified, we get 
\begin{eqnarray}\label{14ab}
 \Big[\Big(\frac{u^r}{u^t}\Big)^2 -a^2\Big]\partial_r(\ln{u^r)} &=& \frac{1}{r(u_t^2)}\Big[2(a u_t)^2\\&-&\frac{(1+\alpha)M}{r}\big(1-\frac{\alpha M}{r}\big)\Big].
\end{eqnarray}
As at sonic point $a_*^2 = (\frac{u^r}{u_t})^2$, so we obtain:
\begin{eqnarray}\label{15ab}
2a_{*}^{2}(u_{t_*})^2&=& \frac{(1+\alpha)M}{r_*}\big(1-\frac{\alpha M}{r_*}\big).
\end{eqnarray}
After substituting above in (\ref{15}), we have 
\begin{eqnarray}\label{20}
({u^r}_*)^2&=&\frac{(1+\alpha)M}{2r_*}\big(1-\frac{\alpha M}{r_*}\big).
\end{eqnarray}
Using the equations (\ref{13aaa}), (\ref{15ab}) and (\ref{20}), we get he value:
\begin{eqnarray}\label{21}
a_{*}^{2}&=&\frac{{(1+\alpha)(r_*-\alpha M)M}}{2r_*^2-3(1+\alpha)r_*M+\alpha(1+\alpha)M^2}.
\end{eqnarray}
Using Eq. (\ref{20}) and (\ref{21}), we can find the critical points $(r_*,u^r_*)$.

\section{Test fluids}
The isothermal test fluid model represents accreting matter, where the temperature remains constant throughout the flow. This assumption ensures that the speed of sound remains uniform at all radial distances, simplifying the analysis of the accretion process. As the fluid has a high velocity, there is minimal heat exchange with the environment, effectively causing the flow to behave adiabatically despite the isothermal condition. The equation of state (EoS) for such a fluid is expressed as: 
\begin{equation}
    p= k e, 
\end{equation}
where 
$p$ denotes the pressure, $e$ represents the energy density, and $k$ is a dimensionless state parameter constrained to the range $0<k\leq 1$. This proportional relationship between pressure and energy density simplifies the study of accretion dynamics, allowing for analytical and numerical investigation. Specific values of $k$, such as $k=1$ for an ultra-stiff fluid or $k=1/3$ for a radiation-dominated fluid, highlight different physical regimes \cite{t1}. 
The adiabatic sound speed is given by $a^{2}= {dp}/{de}$. By applying the equation of state for isothermal fluids, it can be determined that $a^{2}=k$. Since the process is isentropic, the entropy remains constant, leading to $T dS = 0$, where \( S \) denotes the entropy. This condition aligns with the first law of thermodynamics, the change in energy density with respect to particle number density is given by
\begin{equation} \label{39}
\frac{de}{dn} = \frac{e + p}{n} = h,
\end{equation}
where $h$ is the specific enthalpy.
Integrating this relation yields
\begin{equation} \label{39a}
n = n_* \left( \frac{e}{e_*} \right)^{\frac{1}{k + 1}},
\end{equation}
where $n_*$ and $e_*$  are the particle number density and energy density at the sonic point, respectively.
Comparing this to enthalpy, we get:
\begin{eqnarray}\label{39b}
    h=\frac{(k+1)e_*}{n_*}\Big(\frac nn_*\Big)^{k}.
\end{eqnarray}
Using (\ref{39a}) in (\ref{13aa}) we get:
\begin{eqnarray}\label{39c}
    n^k\sqrt{f(r)+(u^r)^2}=c_3.
\end{eqnarray}
Comparing (\ref{39c}) and (\ref{13}), we Get
\begin{eqnarray}\label{39d}
    \sqrt{f(r)+(u^r)^2}=c_3r^{2k}(u^r)^k.
\end{eqnarray}
Also for isothermal Equation, Eqs (\ref{20}) and (\ref{21}) lead to
\begin{eqnarray}\label{44a}
(u_*^{r})^2&=&\frac{1}{4}r_* f_{*,r_*}=k\Big( \frac{1}{4} r_* f_{*,r_*} + f_{*}\Big).
\end{eqnarray}
and the Hamiltonian for the accretion flow can be defined as \cite{A1,A2,A3}
\begin{equation} \label{41b}
\mathcal{H} = \frac{f^{1-k}}{v^{2k} r^{4k}(1 - v^2)^{1-k}  },
\end{equation}
where $ v$ represents the three-velocity of the radial flow in the equatorial plane, and is given by $v\equiv \frac{dr}{fdt}$. Hence,
\begin{eqnarray}\label{41bb}
v^{2}=\Big(\frac{u}{fu^{t}}\Big)^{2}=\frac{u^{2}}{u_{t}^{2}}=\frac{u^{2}}{f+u^{2}},
\end{eqnarray}
where $u^{r}=u=\frac{dr}{d\tau}$, $u^{t}=\frac{dt}{d\tau}$ and the time component of the covariant four-velocity is $u_{t}=-fu^{t}$.

To extend the analysis from a single point to a continuous flow of fluids, we now examine the behavior of the fluid under different physical conditions by varying the state parameter \( k \). Specifically, we consider several representative cases: \( k = 1 \) for Ultra-Stiff Fluid (USF), \( k = \frac{1}{2} \) for Ultra-Relativistic Fluid (URF), \( k = \frac{1}{3} \) for Radiation Fluid (RF), and \( k = \frac{1}{4} \) for Sub-Relativistic Fluid (SRF).

\subsection{Ultra-Stiff Fluid $(k=1)$}
An USF describes a case where the pressure is equal to the energy density, corresponding to an equation of state parameter \( k = 1 \) and Eq. (\ref{39d}) turns out to be:
\begin{eqnarray}\label{46k1}
    \sqrt{f(r)+(u^r)^2}=c_4r^{2}(u^r),
\end{eqnarray}
simplifying above, we get
\begin{eqnarray}\label{46k1a}
(u^{r})^2&=&\frac{f(r)}{c_4r^4-1}.
\end{eqnarray}
From Eq. (\ref{44a}) we find that $f_{\ast}=0$ and hence $r_{\ast}=r_{+}=r_{h}$. The Hamiltonian in this case will take the form as given by
\begin{eqnarray}\label{46ab}
\mathcal{H} &=& \frac{1}{(vr^{2})^{2}}.
\end{eqnarray}

\begin{figure}[!ht]
\centering
\includegraphics[width=8cm]{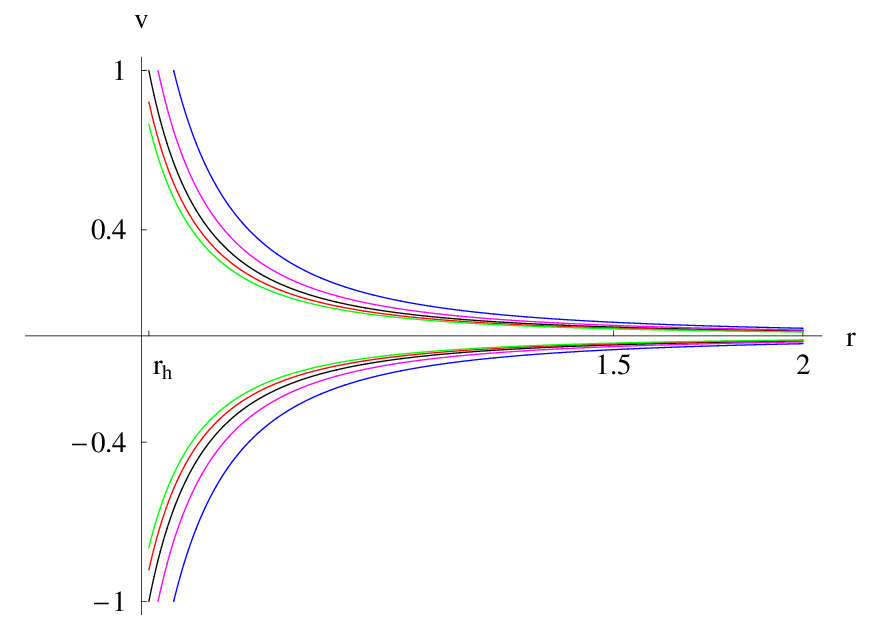}
\caption{Phase space trajectories for different values of the Hamiltonian \( \mathcal{H} \). The black curves indicate the solutions at the critical point where \( \mathcal{H} = \mathcal{H}_{c} \). The green and red curves show solutions with higher Hamiltonian values, specifically \( \mathcal{H} = \mathcal{H}_{\ast} + 50.004 \) and \( \mathcal{H} = \mathcal{H}_{\ast} + 100.007 \), respectively. In contrast, the magenta and blue curves represent lower values, with \( \mathcal{H} = \mathcal{H}_{\ast} - 50.004 \) and \( \mathcal{H} = \mathcal{H}_{\ast} - 100.007 \).}\label{f1}
\end{figure}

We have plotted Eq.(\ref{46ab}) in phase space, with the parameters chosen as \( k = 1 \), \( M = 1 \), and \( \alpha = 0.5 \). The plot in figure (\ref{f1}) illustrates the trajectories of solutions. The black curves indicate the solutions at the critical point where \( \mathcal{H} = \mathcal{H}_{c} \). The red and green curves show solutions with higher Hamiltonian values, specifically \( \mathcal{H} = \mathcal{H}_{\ast} + 50.004 \) and \( \mathcal{H} = \mathcal{H}_{\ast} + 100.007 \), respectively. In contrast, the magenta and blue curves represent lower values, with \( \mathcal{H} = \mathcal{H}_{\ast} - 50.004 \) and \( \mathcal{H} = \mathcal{H}_{\ast} - 100.007 \). The plot illustrates two branches of fluid motion: the upper region corresponds to supersonic accretion flows (\( v > 0 \)), while the lower region represents subsonic flows (\( v < 0 \)).

\subsection{Ultra-Relativistic fluid $(k=1/2)$}
An URF refers to a theoretical fluid where the particles are moving at speeds very close to the $c$ i.e speed of light. This type of fluid is described by a specific EoS as:  $p=e/2$. So it turns out that in an URF the energy density is half of the pressure and Eq. (\ref{39d}) turns out to be:
\begin{eqnarray}\label{46k12}
    \sqrt{f(r)+(u^r)^2}=c_4r(u^r)^{1/2},
\end{eqnarray}
simplifying above, we get
\begin{eqnarray}\label{46k12a}
u^{r}&=&c_4r^2\pm\frac12\sqrt{c_5r^4-4f(r)}.
\end{eqnarray} The expressions for critical point can be find by the expression $r_* f_{*,r_*}-4f_*=0$. Hence,
\begin{eqnarray}\label{50ab}
r_{*}&=&\frac{10M^2\alpha(1+\alpha)4 r_*^2}{\big[18M(1+\alpha\big]},
\end{eqnarray}
and from Eq. (\ref{41b})
 \begin{eqnarray}\label{51ab}
\mathcal{H} &=& \frac{f^\frac12}{(1-v^{2})^{\frac12}vr^2}
\end{eqnarray}
\begin{figure}[!ht]
\centering
\includegraphics[width=6cm]{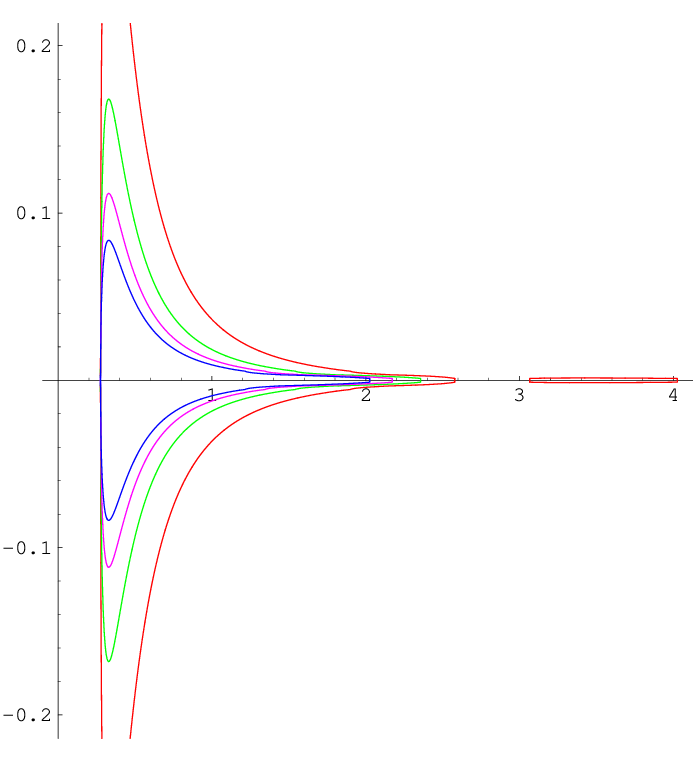}
\caption{Phase space trajectories of the solutions to Eq.~(\ref{51ab}) for \( \alpha = 0.5 \), \( M = 1 \), and \( k = 1/2 \). The black curve represents the critical solution (\( \mathcal{H} = \mathcal{H}_c \)), while the green, magenta, red and blue curves correspond to increasing values of the Hamiltonian: \( \mathcal{H}_\ast + 30.04 \), \( \mathcal{H}_\ast + 60.09 \), \( \mathcal{H}_\ast + 90.04 \), and \( \mathcal{H}_\ast + 120.09 \), respectively. Supersonic motion appears in the region \( v > 0 \), and subsonic motion in \( v < 0 \). Vertical lines highlight regions of unphysical fluid behavior.}
\label{f2}
\end{figure}
Figure (\ref{f2}) displays the phase space trajectories of the solutions to Eq.~(\ref{51ab}) for \( \alpha = 0.5 \), \( k = 1/2 \), and \( M = 1 \). The black curve represents the solution at the critical point where \( \mathcal{H} = \mathcal{H}_{c} \). The green, magenta, red, and blue curves correspond to Hamiltonian values of \( \mathcal{H}_{\ast} + 30.04 \), \( \mathcal{H}_{\ast} + 60.09 \), \( \mathcal{H}_{\ast} + 90.04 \), and \( \mathcal{H}_{\ast} + 120.09 \), respectively. The plot shows two distinct regions of motion: supersonic flow occurs where \( v > 0 \), and subsonic flow occurs where \( v < 0 \). Vertical lines indicate regions where the fluid exhibits unphysical behavior.

\subsection{Radiation Fluid $(k=1/3)$}
A RF is a theoretical model of a fluid in which the energy density is dominated by radiation, such as photons or other massless particles. It is particularly relevant in astrophysics and cosmology for describing the behavior of matter in the early universe or in highly energetic astrophysical environments. For radiation fluid EOS will be $p=e/3$. 
So it turns out that in a RF the energy density is three times of the pressure and Eq. (\ref{39d}) turns out to be:
\begin{eqnarray}\label{46k13}
    \sqrt{f(r)+(u^r)^2}=c_3r^{2/3}(u^r)^{1/3},
\end{eqnarray}
simplifying above, we get
\begin{eqnarray}\label{46k13a}
(u^{r})^2&=&\frac{\Big(f(r)+(u^{r})^2\Big)^3}{c_4^6\ r^4}.
\end{eqnarray} 
The expression to find the critical point will be $r_* f_{*,r_*}-2f_*=0$. Hence,
\begin{eqnarray}\label{54ab}
2r_{*}^2-6M(1+\alpha)r+4M^2\alpha(1+\alpha)&=&0
\end{eqnarray}
and from Eq. (\ref{41b})
 \begin{eqnarray}\label{55ab}
\mathcal{H} &=& \frac{f^\frac23}{(1-v^{2})^{\frac23}v^{\frac23}r^\frac43}
\end{eqnarray}
\begin{figure}[!ht]
\centering
\includegraphics[width=6cm]{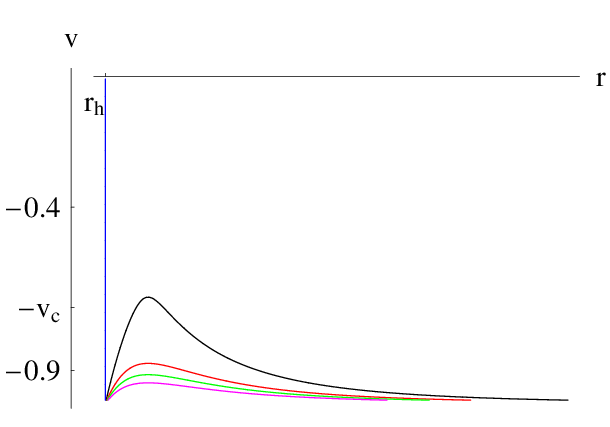}
\caption{Phase space trajectories of the solutions to Eq.~(\ref{55ab}) for \( \alpha = 0.5 \), \( M = 1 \), and \( k = 1/3 \). The black curve corresponds to the solution near the critical point with \( \mathcal{H} = \mathcal{H}_{c} + 10.04 \), while the red, green, magenta, and blue curves represent solutions for \( \mathcal{H} = \mathcal{H}_{\ast} + 15.04 \), \( \mathcal{H} = \mathcal{H}_{\ast} + 18.09 \), \( \mathcal{H} = \mathcal{H}_{\ast} + 22.04 \), and \( \mathcal{H} = \mathcal{H}_{\ast} \), respectively. }\label{f3}
\end{figure}
The phase space trajectories of the solutions to Eq.~(\ref{55ab}) are shown in Figure (\ref{f3}) for the parameters \( k = 1/3 \), \( M = 1 \), and \( \alpha = 0.5 \). The black curve corresponds to the solution near the critical point, with \( \mathcal{H} = \mathcal{H}_{c} + 10.04 \). The  green, red, magenta, and blue curves represent solutions with Hamiltonian values \( \mathcal{H} = \mathcal{H}_{\ast} + 15.04 \), \( \mathcal{H} = \mathcal{H}_{\ast} + 18.09 \), \( \mathcal{H} = \mathcal{H}_{\ast} + 22.04 \), and \( \mathcal{H} = \mathcal{H}_{\ast} \), respectively. As seen in Figure (\ref{f3}), the fluid velocity remains below the local speed of sound, indicating that all trajectories in this case correspond to subsonic flow.
\subsection{Sub-Relativistic Fluid $(k=1/4)$}
A SRF represents a theoretical model with a specific proportionality between pressure and energy density. While not commonly tied to specific physical substances, it serves as a useful approximation in theoretical studies of fluid dynamics, particularly in astrophysical or cosmological scenarios where sub-relativistic velocities and moderate compressibility are relevant. For SRF, equation of state will be $p=e/4$. 
So it turns out that the energy density is quarter of the pressure and Eq. (\ref{39d}) become:
\begin{eqnarray}\label{46k14}
    \sqrt{f(r)+(u^r)^2}=c_3r(u^r)^{1/4},
\end{eqnarray}
simplifying above, we get
\begin{eqnarray}\label{46k14a}
(u^{r})^2&=&\frac{\Big(f(r)+(u^{r})^2\Big)^2}{c_4^2\ r^2}.
\end{eqnarray} 
The expression for finding critical point will be $3 r_* f_{*,r_*}-4f_*=0$. Hence,
\begin{eqnarray}\label{58ab}
2r_{*}^2-7M(1+\alpha)r+5M^2\alpha(1+\alpha)&=&0.
\end{eqnarray}
and from Eq. (\ref{41b})
\begin{eqnarray}\label{59ab}
\mathcal{H} &=& \frac{f^\frac34}{(1-v^{2})^{\frac34}v^\frac12}r,
\end{eqnarray}
\begin{figure}[!ht]
\centering
\includegraphics[width=6cm]{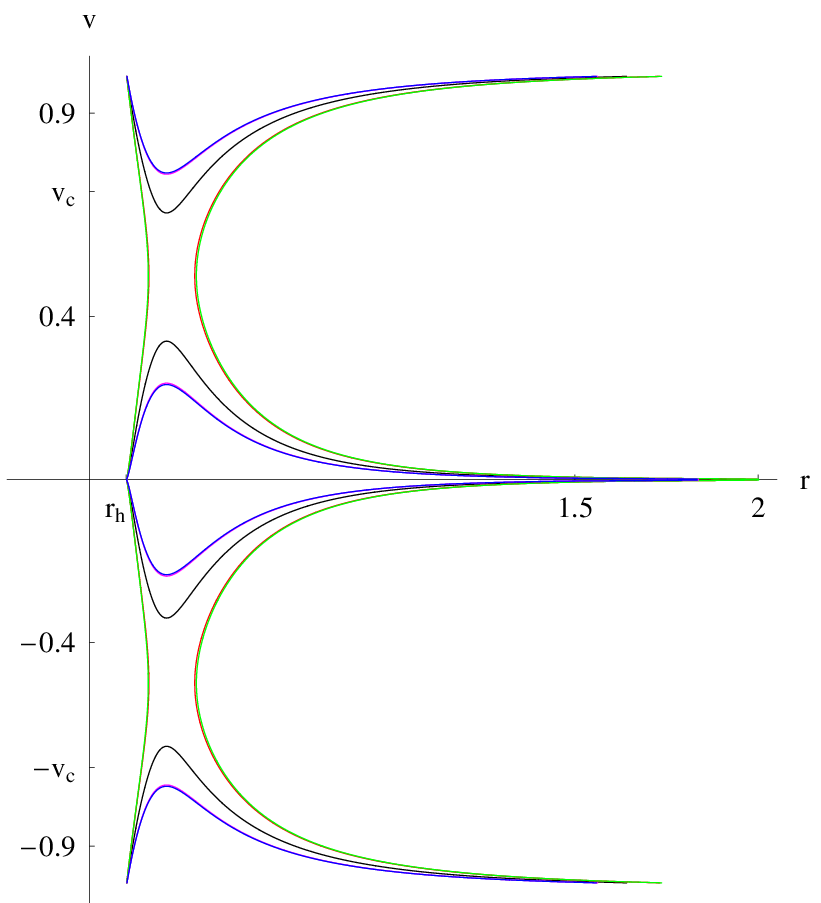}
\caption{Phase trajectories corresponding to solutions of Eq.~(\ref{59ab}) for \( \alpha = 0.5 \), \( M = 1 \), and \( k = 1/4 \). The black curve shows the critical solution where \( \mathcal{H} = \mathcal{H}_c \), while the red and green curves represent lower Hamiltonian values (\( \mathcal{H}_\ast - 1.04 \) and \( \mathcal{H}_\ast - 1.09 \)). The magenta and blue curves correspond to higher values (\( \mathcal{H}_\ast + 1.04 \) and \( \mathcal{H}_\ast + 1.09 \)). The figure highlights the transition between supersonic (\( v > v_c \)) and subsonic (\( v < v_c \)) flow regimes.}\label{f4}
\end{figure}
Figure (~\ref{f4}) presents the phase space trajectories of the solutions to Eq.~(\ref{59ab}), with parameters set to \( k = 1/4 \), \( M = 1 \), and \( \alpha = 0.5 \). The black curve corresponds to the critical solution where \( \mathcal{H} = \mathcal{H}_c \). The green and red curves represent solutions with slightly lower Hamiltonian values, \( \mathcal{H} = \mathcal{H}_\ast - 1.04 \) and \( \mathcal{H} = \mathcal{H}_\ast - 1.09 \), respectively. In contrast, the magenta and blue curves correspond to higher values, \( \mathcal{H} = \mathcal{H}_\ast + 1.04 \) and \( \mathcal{H} = \mathcal{H}_\ast + 1.09 \). The plot illustrates a clear distinction between two flow regimes: supersonic flow occurs when \( v > v_c \), while the region where \( v < v_c \) represents subsonic motion.

\section{Black hole's Accretion Rate}
The mass accretion rate of a BH, denoted by 
$\dot{M}$ represents the rate at which the BH accumulates mass from its surroundings. It is a fundamental parameter in BH astrophysics, quantifying the amount of matter accreted per unit time. Mathematically, the accretion rate is defined as the product of the event horizon area and the flux of matter crossing it. This process drives the growth of BHs and significantly impacts their interaction with the surrounding environment. As matter falls toward a BH, it generally forms an accretion disk, where viscous and frictional forces facilitate angular momentum loss, allowing the material to spiral inward. The gravitational energy of the infalling matter is converted into heat, producing intense radiation across multiple wavelengths. In this section explores how the accretion rate changes with radial distance. The expression used to determine the rate of accretion is: \cite{A3}
\begin{eqnarray}\label{61-2}
\dot{M} &=& 4\pi A_1 (e + p).
\end{eqnarray}
For isothermal fluids $p=ke$, which implies that $(e+p)=e(1+k)$. Hence,
\begin{eqnarray}\label{61-6}
e&=&\frac{\left( A_0^2 r^4 - 4 A_1^4 \right)}{4 A_1^2 r^4 f(r) }.
\end{eqnarray}
Putting (\ref{61-6}) into (\ref{61-2}), we get
\begin{eqnarray}\label{61-7}
\dot{M}&=&\frac{ 2 \pi  \left( A_0^2 r^4 - 4 A_1^4  \right)}{ A_1 r^2 \left[ r^{2}-2M(1+\alpha)r+M^{2}\alpha(1+\alpha)  \right]}.~~~~~
\end{eqnarray}
\begin{figure}[!ht]
\centering
\includegraphics[width=8cm]{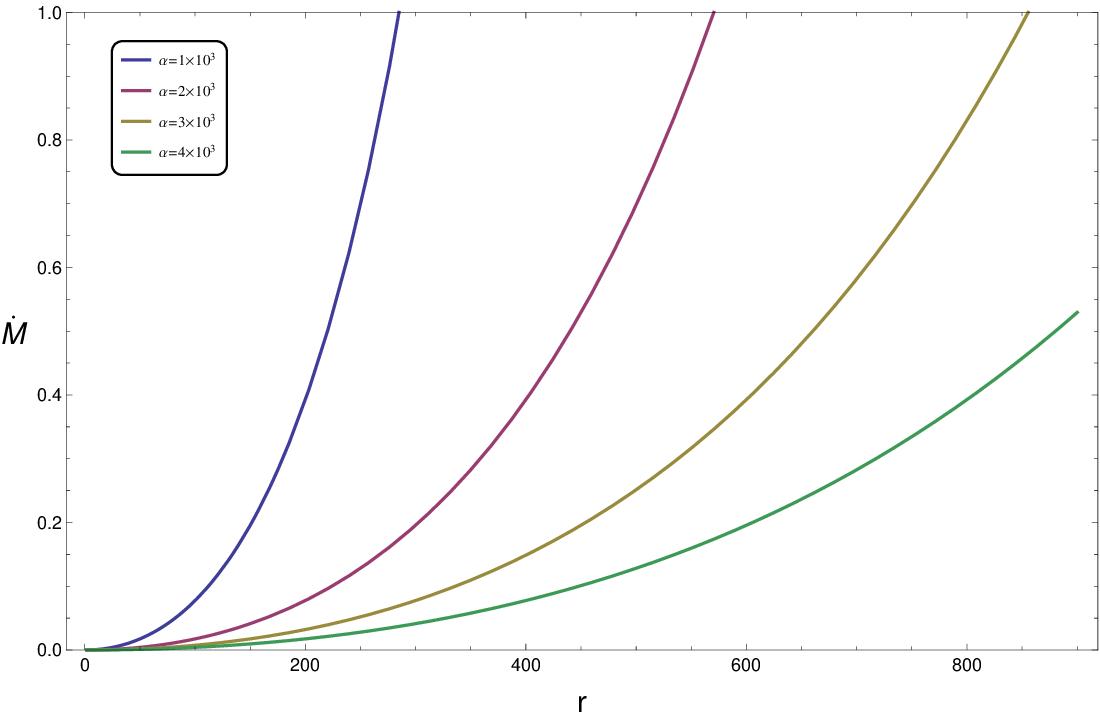}
\caption{The plot illustrates the variation in mass accretion rate for different values of the MOG parameter \( \alpha \), considering an USF characterized by \( k = 1 \). The other parameters are fixed and set to \( M = A_0 = A_1 = 1 \). The plot demonstrates that the accretion rate increases with both the radial distance and the MOG parameter, indicating the enhanced influence of modified gravity on the accretion process.} \label{f5}
\end{figure}

In the figure ~\ref{f5} shows the variation of the mass accretion rate for different values of the MOG parameter \(\alpha\), chosen from within the range 
\(-1 \leq \alpha \leq \infty \). The plot illustrates how the accretion rate changes as \(\alpha\) varies. It is observed that the accretion rate increases with the radial distance $r$ and larger values of the MOG parameter 
\(\alpha\) also lead to a higher accretion rate.
\section{Conclusion}
Accretion onto BHs remains a key area of interest in theoretical astrophysics, particularly in the context of modified gravity theories. In this work, we have investigated the process of spherical accretion onto Schwarzschild–MOG (SH-MOG) BHs. The analysis began with the removal of coordinate singularities, followed by the formulation of the general accretion framework and identification of the sonic point. Notably, by setting the MOG parameter \( \alpha = 0 \), we recovered the classical Schwarzschild accretion scenario, consistent with general relativity.

We explored the accretion dynamics of isothermal test fluids by varying the equation of state parameter \( k \), which characterizes different fluid types. For \( k = 1 \) (USF), the phase space reveals two distinct branches corresponding to supersonic (\( v > 0 \)) and subsonic (\( v < 0 \)) flows. Similar behavior was observed for \( k = 1/2 \) (URF), although some trajectories exhibit unphysical features marked by vertical boundaries. For \( k = 1/3 \) (RF), the flow remains entirely subsonic. In the case of \( k = 1/4 \) SRF), the flow once again separates into supersonic and subsonic regimes, divided by a critical velocity \( v_c \).

Furthermore, we discussed the influence of the MOG parameter \( \alpha \) on the mass accretion rate. Our findings indicate that the accretion rate increases with both the radial distance \( r \) and the value of \( \alpha \), suggesting that the effects of modified gravity enhance the inflow of matter onto BHs.

\section{Declaration of Competing Interest}
The authors declare no competing financial or personal interests that could have influenced the research presented in this paper. This study is conducted solely for academic and scientific purposes, with no external funding or conflicts of interest affecting the results and conclusions.

\section{Data availability}
No data was used for the research described in the article.

\end{document}